\documentclass[letter,traditabstract]{aa}  
\usepackage{graphicx}
\usepackage{txfonts}
\usepackage{natbib}
\newcommand{\gilFull}{GBT 1355+5439}
\newcommand{\gil}{GBT 1355}
\newcommand{\HI}{\ion{H}{i}}
\newcommand{\kms}{km\ s$^{-1}$}
\newcommand{\mJybeam}{mJy\ beam$^{-1}$}
\newcommand{\Msol}{$M_\odot$}

\begin{document} 


\title{Is \gilFull\ a dark galaxy?}

\author{T.A. Oosterloo\inst{1,2}
        \and
          G.H. Heald\inst{1}
        \and 
          W.J.G. de Blok\inst{1,3}
}

\institute{Netherlands Institute for Radio Astronomy (ASTRON), Postbus 2, 
           7990AA Dwingeloo, The Netherlands
\and
           Kapteyn Astronomical Institute, University of Groningen, 
           Postbus 800, 9700AA, Groningen, The Netherlands
\and
           Astrophysics, Cosmology and Gravity Centre (ACGC), 
           Department of Astronomy, University of Cape Town, 
           Private Bag X3, Rondebosch 7701, South Africa
}

\date{Received 26 May 2013 / Accepted 14 June 2013 }
 
\abstract{We present \HI\ imaging of \gilFull\ performed with the Westerbork Synthesis Radio Telescope. This is a dark \HI\ object recently discovered  close to the nearby galaxy M101. We find \gilFull\ to be an \HI\ cloud $5\times3$ arcmin in size. The total \HI\ image and the kinematics show that the cloud consists of condensations that have small ($\sim$10 \kms) motions with respect to each other. The column densities of the \HI\ are low; the observed peak value is $7.1\times10^{19}$ cm$^{-2}$. The velocity field shows a mild velocity gradient over the body of \gilFull, possibly due to rotation, but it may also indicate large-scale radial motions. Although our data are limited in sensitivity, at all positions the \HI\ velocity dispersion is higher than 5 \kms\ and no narrow, cold,  \HI\ component is seen. 
Because its distance is not known, we considered various possibilities for the nature of \gilFull.  Both the scenarios that it is a tidal remnant near M101 and that it is a dark dwarf companion of M101 meet difficulties. Neither do the data fit the properties of known compact High-Velocity Clouds in the Galactic halo exactly, but we cannot entirely exclude this option and deeper observations are required. We also considered the possibility that \gilFull\ is a gas-rich dark minihalo in the outer regions of the Local Group. Interestingly, it would then have similar properties as the clouds of a proposed Local Group population  recently found in the ALFALFA survey. In this case, the \HI\ mass of \gilFull\ would be about a few times $10^5$ \Msol,  its size about 1 kpc, and the dynamical mass $M_{\rm dyn} > 5\ \times\ 10^7$ \Msol. However, if \gilFull\ is a dark Local Group object, the internal kinematics of the \HI\  appears to be different from that of gas-dominated, almost dark galaxies of similar size.
}


\maketitle


\section{Introduction}

One of the puzzles of extragalactic astronomy is the formation of the smallest galaxies. Because of their shallow dark matter potentials, they are very delicate and sensitive systems, and modelling their  evolution requires an accurate description and balance of all physics. A number of years ago, there appeared to be a large discrepancy between predictions and observations regarding the number of dwarf galaxies \citep{Klypin1999,Moore1999}. However, due to a large amount of theoretical work and new optical surveys \citep[see][]{McConnechie2012},  the gap between observations and theory has considerably decreased. 

Nevertheless, some important questions remain. A key feature of galaxy formation is that star formation becomes increasingly inefficient for progressively smaller galaxies. This leads to the question whether there is a minimum halo mass below which no stars form, and, if there were such a limit, whether there are starless objects, but that are nevertheless visible in other tracers, e.g.\ through their cold gas \citep[e.g.][]{Sternberg2002,Maloney2003,Bovill2011}. Several searches for such gas-rich 'dark galaxies' have been made. Although a number of objects have been found that are almost dark,  there is no observational evidence for an extensive population of gas-rich dark objects, either in the Local Group or in other nearby systems \citep[][but see \citet{Giovanelli2010} for an interesting new possibility]{Zwaan2000,Meyer2004,Pisano2007,Chynoweth2011}.

Recently, a deep survey of a region around the nearby galaxy M101, conducted in single-dish \HI\ (GBT) and in the optical,    revealed  an, at the GBT resolution of 9 arcmin, unresolved isolated  \HI\  cloud (\gilFull, hereafter \gil) near M101. It  has no optical counterpart, even in very deep optical images \citep{Mihos2012,Mihos2013}. The immediate question is, of course, whether the object is a dark galaxy. Answering this question requires spatially resolved \HI\ data, and here we report on interferometric observations that have allowed us to image the distribution and kinematics of the object.  Although the detection of the \HI\ in \gil\ is fairly faint, a number of interesting features are detected that may give some clues to the nature of this object.

\begin{figure}
  \centering
  \includegraphics[width=7.5cm]{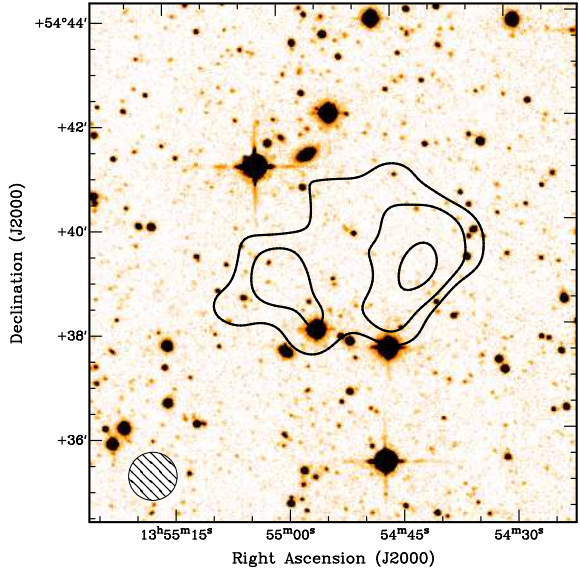}
  \caption{Integrated \HI\ contours on the deep $V$-band optical image from \citet{Mihos2013}, kindly provided to us by Chris Mihos. The 1-$\sigma$ limiting surface brightness in this optical image is $\mu_V = 29$ mag arcsec$^{-2}$. \HI\ contour levels are 1.5, 3 and $6 \cdot 10^{19} $ cm$^{-2}$. 
  }
  \label{FigColDens}%
\end{figure}


\section{Data}

The neutral hydrogen in \gil\ was imaged using a standard 12-hr observation with the WSRT on January 29, 2013. A 10 MHz bandwidth was used, divided into 2048 channels, giving a nominal velocity resolution of about 1 \kms.  Standard WSRT calibration and data reduction procedures were applied  using the MIRIAD package \citep{Sault95}.

{ The initial inspection of the data immediately made clear that the \HI\ column densities in \gil\ are below $10^{20}$ cm$^{-2}$: in data cubes made with  standard full spatial resolution only a hint of \HI\ is visible. } Only when the data are tapered to lower spatial and velocity resolution, a clear detection of \gil\ is made. A good compromise between resolution and signal-to-noise ($S/N$) was found  by smoothing the data spatially to a resolution of 55 arcsec.

To construct the total \HI\ image, the data were binned to a velocity resolution  of 16.4 \kms. This was done so that the \HI\  is found in only 1-2 channels, maximising the $S/N$ for detection.  
The noise in this data cube  is 0.5 \mJybeam\ and the 5-$\sigma$ detection limit per 3-D resolution element is $1.45\times10^{19}$ cm$^{-2}$. The total \HI\ image obtained in this way is shown in Fig.\ \ref{FigColDens}.

The overall shape of \gil\ is roughly elliptical, oriented east-west, extending $5\times3$ arcmin. The \HI\ distribution shows two concentrations with fainter \HI\ in between. The observed total \HI\ flux  is $1.05 \pm 0.15$ Jy \kms. This is basically the same  as derived by \citet{Mihos2012} from the GBT data, that have  a spatial resolution of 9\farcm1 (roughly the size of Fig.\ \ref{FigColDens}). This indicates that all the \HI\  detected with the GBT  is represented in Fig.\ \ref{FigColDens}. 

Figure  \ref{FigColDens} confirms that the \HI\ column densities are indeed low: the peak value is $7.1\times10^{19}$ cm$^{-2}$ at the western side of the object; on the eastern side a second peak is observed with a column density of $4.9\times10^{19}$ cm$^{-2}$. The range of column densities observed is lower than the critical column density below which no star formation is expected to occur \citep[e.g.,][]{Schaye2004}.

\begin{figure}
  \centering
  \includegraphics[width=7.5cm]{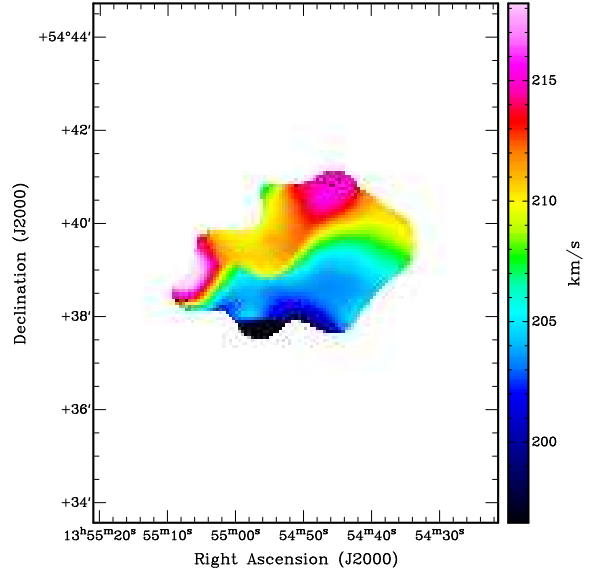}
  \caption{Velocity (heliocentric) field of the \HI\ in \gil.}
  \label{FigVel}%
\end{figure}

The kinematics of the \HI\ was studied by fitting Gaussians to the spectra of a data cube made with the same spatial tapering as used above, but now with a velocity resolution of 4.2 \kms. The velocity field  is shown in Fig.\ \ref{FigVel}. This figure shows that there is a  N-S velocity gradient of about 15 \kms\ over the object, more or less perpendicular to the morphological major axis.  This velocity gradient could indicate overall rotation of the \HI, but given that the gradient is aligned with the minor axis of the \HI\ body, it may  as well indicate a radial flow (in or out) of the gas. Below we show that at some locations the spectra show multiple components. This means that the velocity field only crudely represents the overall kinematics. The heliocentric systemic velocity  was found to be $V_{\rm sys} = 208\pm 5$ \kms, consistent with the value of 210 \kms\ found by \citet{Mihos2012}.

\begin{figure}
  \centering
  \includegraphics[width=7.5cm]{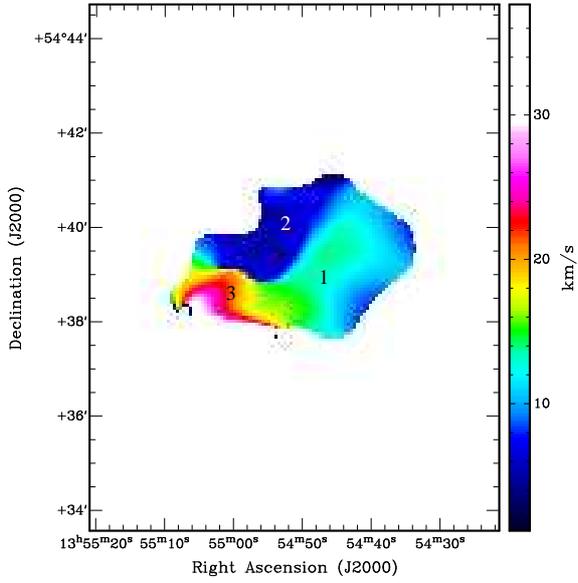}
  \caption{Velocity dispersion of the \HI\ in \gil. The numbers indicate where the spectra given in Fig.\ \ref{FigProf} were taken.}
  \label{FigSig}%
\end{figure}

Figure \ref{FigSig} gives the velocity dispersion map. In the north the lowest dispersions are found, with values between 5 and 8 \kms, while in the south dispersions have values above 20 \kms. In the western part  the dispersions are intermediate with values around 15 \kms. Nowhere  dispersions below 5 \kms\ are  detected, even though the spectral resolution would allow  detecting dispersions as low at 1.5 \kms.

\begin{figure}[b]
  \centering
  \includegraphics[width=8cm]{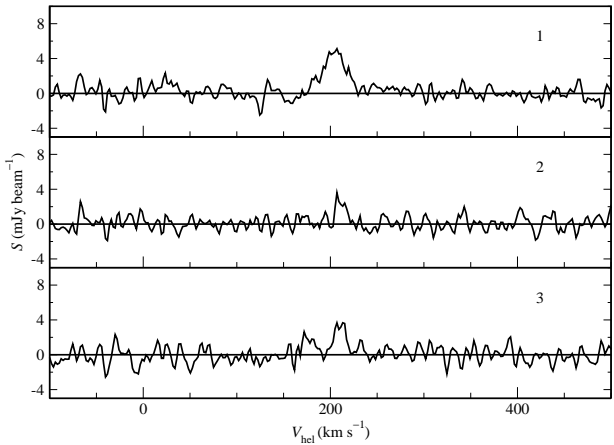}
  \caption{Spectra taken at the positions indicated in Fig.\ \ref{FigSig}. 
  }
  \label{FigProf}%
\end{figure}

There is quite some structure in the spatial distribution of the dispersions, and there appears to be a fairly sharp transition between the relatively narrow profiles in the north and the wide profiles in the south. There are indications that the high dispersions in the south are due to fitting a single Gaussian to profiles that are a sum of two narrower components. This is illustrated in Fig.\ \ref{FigProf} where three spectra are shown. Spectrum 1 is taken in the western peak of the \HI\ distribution where the intermediate dispersions are found, while spectrum 2 is taken in the region of lower velocity dispersion. Spectrum 3 is taken from the region with highest dispersions and appears to be composed of two narrower components. The most likely explanation  is that \gil\ contains a small number of gas concentrations that have small random motions with respect to each other and  that in the southern part we see two such concentrations superposed.

\section{Nature of \gil}

In the following, we discuss a number of possibilities for the nature of \gil. One obvious scenario is that it is an object in the vicinity of M101. However, 
because only one dark \HI\ cloud is detected near M101, while the sensitivity of the data is good enough to have detected much fainter clouds as well, \gil\ may have no physical relation to M101 and the proximity to M101 on the sky could be a chance superposition. Therefore, we also discuss  the possibilities that \gil\ is a foreground object in the Galactic halo or in the Local Group.

\subsection{Object near M101}

First we discuss the possibility that \gil\ is an object associated with M101. If this is the case\footnote{We assume the same distance to M101 as in \citet{Mihos2012}:  6.9 Mpc. For this distance, 1 arcmin corresponds to 2.0 kpc.}, its size would be $10 \times 6$ kpc and the \HI\ mass $1.1\ \times\ 10^7$ \Msol. The projected distance to the centre M101 would be 150 kpc.


One possibility is that \gil\ is a remnant of a tidal interaction with one of the members of the M101 group. Such remnants are seen near many large galaxies \citep[see e.g.,][]{Sancisi2008}. \citet{Mihos2012} discussed this possibility and concluded that there is no observational evidence that \gil\ is associated with an ongoing or recent tidal interaction. While \gil\ is detected with high $S/N$ in their data, they did not detect any other \HI\ that could hint at a tidal nature of \gil. In addition, in the deep optical images of \citet{Mihos2013} no indications for such an interaction were found either.

We can add to this that the shape and kinematics do not quite match that what is generally expected for tidal filaments.  Because of the nature of tidal forces, tidal filaments are usually  linear structures, different from \gil. Moreover, velocity gradients in tidal filaments tend to be aligned with such filaments, while the velocity gradient in \gil\ is perpendicular. 

If \gil\ is not a tidal filament, it could be a 'proper' galaxy near M101. However, it would be then have to be  of exceptionally low optical surface brightness. For example, in the compilation of dwarf galaxies in and around the Local Group of \citet{McConnechie2012}, all dwarfs detected in \HI\ have central surface brightnesses brighter than  26 mag arcsec$^{-2}$ in $V$, with most  brighter than $\mu_{\circ,V} = 24$ mag arcsec$^{-2}$.  Any optical emission from \gil\ must be much fainter:  the upper limit to the $V$ surface brightness is 29 mag arcsec$^{-2}$ \citep{Mihos2013}. 

Assuming \gil\ is a gravitationally bound object,  we can  estimate  its total dynamical mass. One uncertainty is that we do not know the shape  of the total matter distribution, and therefore do not know exactly how to use the kinematical information. A frequently used  compromise   is  \citep{Hoffman1996}
\[
M_{\rm dyn} = 2.325 \times 10^5  \ 
             \left({V^2_{\rm rot} + 3\sigma^2 \over {\rm km}^2\ {\rm s}^2 }\right)
             \left({ r \over {\rm kpc}}\right)\ 
             M_\odot.
\]

Another complication is  that one does not know whether to interpret the velocity gradient as rotation, and even if one could do so,  the inclination would not be well constrained. We therefore derived a lower limit to the dynamical mass by assuming that there is no systematic rotation. In this way, we found $M_{\rm dyn} > 3.5\times10^8$ \Msol, using $\sigma = 10$ \kms\ and $r = 5$ kpc. Except for the case where \gil\ is seen fairly face on, the effect of possible rotation  is modest. For example, assuming that the velocity gradient indicates rotation and, in addition, the inclination is $30^\circ$, 
we find $M_{\rm dyn} = 6.1\times10^8$ \Msol.

Interestingly, if \gil\ is a dark object near M101, its properties would be very similar to those of the compact High-Velocity Clouds (CHVCs) in the model proposed by \citet{Braun1999}. In this model, the observed CHVCs  are  a population of unevolved  (dark) dwarf galaxies scattered throughout the Local Group, at distances up to 1 Mpc. This model has met with a number of objections, however. One is that such dark objects  would be too large to be consistent with theoretical models of small, starless dark matter halos containing observable \HI\  \citep{Sternberg2002,Maloney2003}. Another objection is that surveys of nearby galaxy groups have not detected  similar populations of similarly sized \HI\ clouds \citep[e.g.][]{Zwaan2000,Pisano2007,Chynoweth2011}.   The same objections would apply here. According to the CHVC model of \citet{Braun1999},  one would expect a substantial population of  failed dwarf galaxies near M101, but the data of \citet{Mihos2012} exclude this.

\subsection{Galactic object}

The above discussion shows that  there could be some problems with the scenario of \gil\ as an object in the direct vicinity of M101. We therefore discuss two other possibilities, the first one being that  \gil\ is an \HI\ cloud associated with our own Galaxy. Given its high Galactic latitude ($60^\circ$)  and its velocity, \gil\ would then be a CHVC somewhere in the Galactic halo. If so, it would be quite a small cloud: for a distance of 10 kpc, the physical size would be  $14.5\ \times\ 8.7$ pc and its \HI\ mass  only $24.7$ \Msol. 

However, if \gil\ is Galactic, it would  be a somewhat unusual CHVC. As \citet{Mihos2012} discussed, within a radius of several tens of degrees from \gil\ there  are indeed a number of known Galactic HVCs, but they are at quite different velocities. Almost all these HVCs have {\sl negative} velocities. The highest velocity of a Galactic HVC in this region of sky is +50 \kms\ \citep[see][]{Wakker1991} therefore the velocity differences between \gil\ and the nearby HVCs are at least 150 \kms, ruling out any association. However, the observations used by \citet{Wakker1991} are not deep enough to have detected small clouds like \gil, so the possibility remains that there is a population of smaller CHVCs in this region of the sky.  \citet{Mihos2012} noted, however, that they found no trace of any other Galactic \HI\ that could be associated with \gil\ in their deep GBT observations.

Some additional information comes from the kinematics. Imaging studies of CHVCs have shown \citep[e.g.][]{deHeij2002,Bruns2004} that they  typically consist of cores  with  line widths of at most a few \kms\ that are embedded in more diffuse emission with velocity widths similar to those seen in \gil.  \gil\ does appear to consist  of gas condensations surrounded by more diffuse gas. However, a distinct difference seems to be that the profiles of these condensations are much broader than those of the cores of CHVCs. On the other hand, we cannot entirely exclude that the sensitivity of our data is not good enough to have detected such narrow components. We have planned more sensitive observations  to investigate this in more detail.

We note that \gil\ is located on the great circle spanned by the Magellanic Stream \citep{Nidever2010}. As seen from the Magellanic Clouds, it is about 90 degrees beyond the tip of the Leading Arm of the Magellanic \HI\ system. Although the extent of the Leading Arm has not been completely determined, this makes an association with the Leading Arm unlikely.

\subsection{Local Group minihalo}

The last hypothesis we discuss is that \gil\ is a foreground object   located in the Local Group. This is, in fact, an interesting possibility. \citet{Giovanelli2010} and \citet{Adams2013} reported the discovery of a number of  ultra-compact HVCs with properties that, although other interpretations cannot be entirely excluded, are consistent with them being isolated, dark minihalos in the outskirts of the Local Group. If they are assumed to be at a distance of 1 Mpc, these clouds have \HI\ masses in the range of  $10^5$ \Msol\ to $10^6$ \Msol, sizes up to a few kpc, and dynamical masses  of $10^7$ - $10^8$ \Msol. Their heliocentric velocities range up to +320 \kms\ and they have linewidths of about 20-30 \kms. They are basically smaller versions, by roughly a factor 10, of the clouds of the CHVC model  of \citet{Braun1999}. However, the clouds as proposed by \citet{Giovanelli2010} and \citet{Adams2013} {\sl are} consistent with the theoretical models of, e.g., \citet{Sternberg2002}. They would also be small enough to have escaped detection in existing searches for dark \HI\ clouds in nearby groups of galaxies.

Assuming \gil\ is at a distance of 1 Mpc, its properties would be similar to those deduced for these clouds. Its \HI\ mass would be $2.4\times\ 10^5$ \Msol,  its size $1.4\ \times\ 0.9$ kpc, and the dynamical mass $M_{\rm dyn} > 5\ \times\ 10^7$ \Msol. The number density of clouds reported by  \citet{Adams2013} is such that one would expect about one such cloud in the survey area of \citet{Mihos2012}. 

One useful comparison is with very small, almost dark, \HI-dominated galaxies such as Leo T \citep{RyanWeber2008} and Leo P \citep{Giovanelli2013}. The size, gas content, and dynamical mass of these galaxies are very similar to that of \gil, if it is assumed to be at 1 Mpc. The two Leo galaxies do have an observable number of stars, but they are completely gas dominated. Their stellar components are very faint and  $>70\%$ of the visible baryons are in \HI. Therefore,  the stellar components are not very relevant for their overall structure. If \gil\ were a dark minihalo, one would expect that it would be very similar kinematically. This is not the case. Both Leo T and Leo P do not have a clumpy \HI\ distribution and show not much structure in the velocity dispersions. This could argue against the minihalo hypothesis, unless \gil\ happens to be a bound pair of minihalos. \HI\ imaging of the Giovanelli and Adams clouds will be useful to investigate this question in greater detail. The radial velocity and sky location are consistent with \gil\ being a member of the loose group containing Leo P.

\section{Summary}

We have imaged the \HI\ in \gilFull, a starless \HI\ cloud near M101 recently discovered by \citet{Mihos2012}. We find it to be a cloud of $5\times3$ arcmin. The distribution and kinematics indicate that it consists of a small number of gas concentrations that have small relative motions with respect to each other. Because the distance to \gil\ is not known, we discussed a number of possibilities for the nature of this cloud. However, none of these provided a perfect explanation. If the object is assumed to be physically associated with M101, we find that the size, structure and kinematics is not what one would expect for either a tidal filament near M101, or for a dark, starless galaxy in the M101 group. Although the data neither exactly fit the properties of known CHVCs in the Galactic halo, we cannot entirely exclude this option. Deeper observations are required for a more detaild study. If the object is part of the Local Group, its size and mass would be  very similar to that of the clouds from the Local Group population of \HI\ objects recently proposed by \citet{Giovanelli2010} and \citet{Adams2013}. In that scenario, it would also have similar size and mass as that  of some of the smallest known  gas-dominated dwarf galaxies in or near the Local Group and it would only differ by the absence of a very small population of stars. However, the internal kinematic of \gil\ seems to be different from such objects.

\begin{acknowledgements}
We thank Chris Mihos for providing the deep optical images of \gilFull\ in digital form. We also thank the anonymous referee for valuable input. The Westerbork Synthesis Radio Telescope is operated by the ASTRON (Netherlands Institute for Radio Astronomy) with support from the Netherlands Foundation for Scientific Research (NWO). WJGdB was supported by the European Commission Grant FP7-PEOPLE-2012-CIG\#333939
 
\end{acknowledgements}

\end{document}